\documentclass{kluwer}    

\usepackage{graphicx}

\newdisplay{guess}{Conjecture}

\begin{document}                                                                                   
\begin{article}
\begin{opening}         
\title{High-Resolution Observations in B1-IRS: ammonia, CCS and water
  masers} 
\author{Itziar \surname{de Gregorio-Monsalvo}}  
\runningauthor{de Gregorio-Monsalvo et al.}
\runningtitle{Molecular environment of B1-IRS}
\institute{Laboratorio de Astrof\'{\i}sica Espacial y F\'{\i}sica
  Fundamental (INTA), Apartado 50727, E-28080 Madrid, Spain\\
National Radio Astronomy Observatory, P.O Box 0,
  Socorro, NM 87801, USA}
\author{Claire \surname{J. Chandler}}
\institute{National Radio Astronomy Observatory, P.O Box 0,
  Socorro, NM 87801, USA}
\author{Jos\'e F.  \surname{G\'omez}}
\institute{Laboratorio de Astrof\'{\i}sica Espacial y F\'{\i}sica
  Fundamental (INTA), Apartado 50727, E-28080 Madrid, Spain}

\author{ Thomas B. H. \surname{Kuiper}}
\institute{Jet Propulsion Laboratory, California Institute of Technology, USA}
\author{Jos\'e M.  \surname{Torrelles}}
\institute{Instituto de Ciencias del Espacio (CSIC) and Institut d'Estudis Espacials de Catalunya, Edifici Nexus, C/Gran Capit\`a 2-4, 
E-08034 Barcelona, Spain}
\author{ Guillem \surname{Anglada}}
\institute{Instituto de Astrof\'{\i}sica de Andaluc\'{\i}a
  (CSIC), Apartado 3004, E-18080 Granada, Spain}

\date{April, 2004}

\begin{abstract}
We present a study of the structure and dynamics of the star forming region B1-IRS (IRAS 03301+3057) using the properties of different molecules at high angular resolution ($\sim$4$''$).  We have used VLA observations of NH$_3$,
CCS, and H$_2$O masers at 1 cm. 
CCS emission shows three clumps around
the central source,  with a velocity gradient
from red to blueshifted velocities towards the protostar, probably due to the interaction with outflowing material. Water maser
emission is elongated in the same direction as a reflection nebula
detected at 2$\mu$m by 2MASS, 
with the maser spots located in a structure of some hundreds of AU from the 
central source, possibly tracing a jet. We propose a new outflow model 
to explain all our observations, consisting of a molecular outflow near the plane of the sky.
Ammonia emission is extended and anticorrelated
with CCS. We have detected for the first time this anticorrelation  
at small scales (1400 AU) in a star forming region. 

\end{abstract}

\end{opening}           

\section{Motivation}

Molecular spectral line observations are a powerful tool for obtaining
information about kinematics, temperature and density in star
forming clouds. The combination of the observations of several
molecules can be used to give a complete information of the environment
of protostars and the stage of their evolution.

CCS molecular lines are very useful for making
studies related to the structure and the physical conditions of the
clouds, since they are not very opaque but
they are intense and abundant in these regions, and its lack of 
hyperfine structure makes CCS a good molecule for dynamical 
studies (Saito et al. 1987, Suzuki et al. 1992). CCS is a high density tracer,
like NH$_{3}$. The spatial distribution of these two molecules can
be used as a clock to date the age of the clouds, since
a spatial anticorrelation between these two molecules has been observed in dark clouds,
and has been interpreted in terms of chemical evolution (Velusamy et al. 1995, Kuiper et al. 1996).

Mass-loss phenomena are common in the earliest stages of stellar
evolution. This activity can be traced by 
water maser emission (Rodr\'{\i}guez et al. 1980). Moreover these masers provide a good 
characterization of the age of low-mass young stellar objects (YSOs), 
since Class 0 sources are the most probable candidates
to harbor this emission (Furuya et al. 2001).
 
In order to diagnose the physical conditions around YSOs, and study
the dynamics and the stage of evolution, we have analyzed the
properties of CCS, NH$_{3}$, and water masers in the cloud surrounding
the far-infrared source B1-IRS (IRAS 03301+3057), a class 0 source
(Hirano et al. 1997) 
located in the B1 molecular cloud.

\section{Observations} 

Simultaneous observations of the J$_{N}$=2$_{1}$-1$_{0}$ transition of
CCS (rest frequency = 22345.388 MHz) and the 6$_{16}$-5$_{23}$ transition of
H$_{2}$O (rest frequency = 22235.080 MHz) were carried out on 2003
April 4 using the Very Large Array (VLA) of the National Radio
Astronomy Observatory\footnote{The NRAO is operated by
  Associated Universities Inc., under cooperative agreement with the National Science
Foundation.} (USA) in its D configuration.

We have also processed VLA archive data of the NH$_{3}$(1,1)
transition (rest frequency = 23694.496 MHz). The observations were
made on 1988 August 13, with the D configuration. 

In addition to all these radio data, we have retrieved a K-band image
from the image data of the Two Micron All Sky Survey
(2MASS), to get a better position for the infrared source. This image
was processed with AIPS, and convolved with a Gaussian of 5$''$ FWHM, to
search for extended emission.

\section{Results}

The most interesting characteristics of the CCS emission are its clumpy spatial
distribution and the observed velocity gradient. There
are three main clumps surrounding the central source, all of them 
are redshifted
with respect to the systemic velocity of the B1 core ($V_{\rm LSR}$ = 6.3 km
s$^{-1}$, Hirano et al. 1997). The molecular gas in all these clumps
shows clear velocity gradients,
 with less redshifted velocities at positions closer to the central
 object (see Fig. 1).

The ammonia emission is very extended ($>2'$) and
clumpy. The general tendency observed consists in a spatial anticorrelation with
respect to the CCS emission, never reported before at such small scales
($\sim$4$''$, 1400 AU at 350 pc). We think that this anticorrelation is related to the
chemical evolution of the source, but we need to continue with this kind of studies in other sources similar to B1-IRS to derive consistent conclusions about the meaning of this distribution.

Our water maser observation shows a cluster of 23 spots, most of them are
redshifted with respect to the main cloud velocity. They form an elongated
structure (see Fig 2), with a length of $\sim$ 1.$''$3 (455 AU at 350 pc).

We have detected a point-like source at 2 $\mu$m from the 2MASS
archive. 
Convolving the K-band image with a Gaussian of 5$''$ FWHM, we have noticed an extended emission elongated south west from the point
source.

\begin{figure}
\centerline{\includegraphics[width=2.5in,angle=-90]{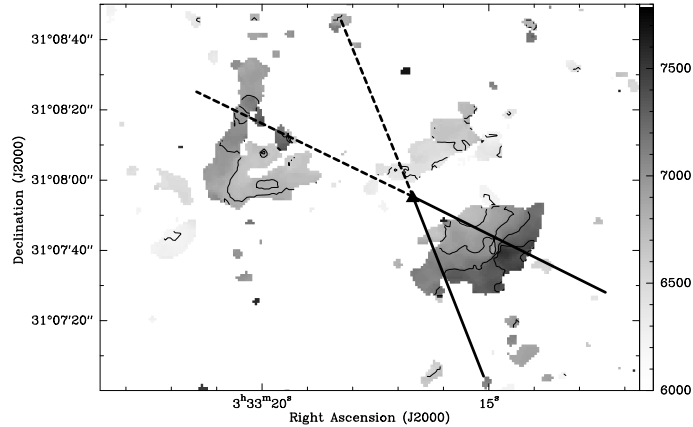}}
\caption{CCS velocity map. Greyscale contours are drawn from 6.0
  10$^{3}$ to 7.8 10$^{3}$ m s$^{-1}$. Levels are represented from 6.0
  10$^{3}$ to 7.8 10$^{3}$
  m s$^{-1}$ by 0.2 10$^{3}$ m s$^{-1}$. The triangle represents the position of the
  2 micron point source. The proposed CO outflow configuration is
  represented by a cone, with solid lines associated with the
  blueshifted CO outflow and the dashed lines associated with the redshifted
  one, with the lobes almost in the plane of the sky.}
\end{figure}

\section{Study of the surrounding environment of B1-IRS}
\subsection{The central source}
The point-like source detected at 2 $\mu$m is located inside the error box of 
the IRAS source (see Fig 3), at a distance of $\simeq$ 6$''$  from the IRAS
catalog position and at $\lsim$1$''$ (350 AU) from the water
masers. Due to their proximity, it is very probable that this 2
$\mu$m source is the exciting source of the maser emission and of the
molecular outflow.

The extended IR emission detected in the  K-band is a reflection nebula
elongated in the same direction as the water masers.

\subsection{The jet-like structure of water masers}
 
Our dynamical calculations reveal that motions of the water masers are  
unbound, since a mass of $\gsim$12 M$_\odot$ within 175 AU would be
needed to bind them. Such a high mass is not consistent with our
NH$_{3}$ and CCS data. Due to  their proximity to the exciting source, and the shape and
size of the distribution of the masers, it is very likely that the maser emission 
delineates the inner (1$''$) part of the jet that is driving the CO outflow,
near B1-IRS and detected by \citeyear{Hir97}.

\begin{figure}
\centerline{\includegraphics[width=3in,angle=-90]{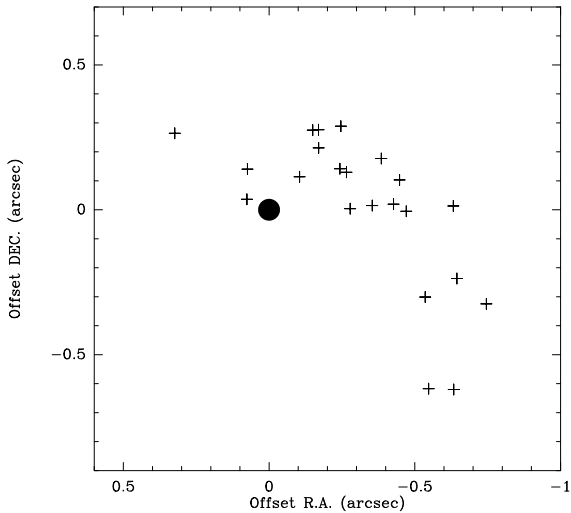}}
\caption{Crosses represent the  water maser components. The
  filled circle represents the position of the 2 micron point source
  (the reference position of this map).}
\end{figure}

\subsection{Interaction between CCS and the molecular outflow}
The velocity pattern observed in the
(redshifted) CCS clumps can be interpreted
in terms of infalling clumps which are strongly interacting with the
  outflow. What we see could be foreground CCS clumps
falling toward the central source, but being stopped by the wind,
explaining in this way the less redshifted velocities observed in the 
CCS lines
at positions closer to the central source (Fig. 1). However, an
alternative possibility is that we could be observing the CCS
clumps in the background moving away from the source and being
accelerated by the molecular outflow.

\section{The geometry of the CO molecular outflow}

The moderately-high velocity blueshifted CO emission  detected by
\citeauthor{Hir97} was proposed to be in a near pole-on
configuration due to the position of the IRAS source with respect to
the outflow, which was located just at the center of the blueshifted
emission (Fig. 3).

Our interferometric data and the results of the 2MASS analysis do not
support that geometry for the outflow, and we propose a new configuration of the outflow material that fits all our results.

First, the 2 $\mu$m point-like source, which is probably powering
the outflow (as suggested by its association with the water masers),
is not located at the center but
at the tip of the blueshifted CO emission.  
The extended 2 $\mu$m emission is elongated in the same direction
that the cluster of water masers, and it is coincident with the shape
of the CO outflow. This coincidence further reinforces that the 2 $\mu$m
source marks the position of the powering source, and that the
blueshifted lobe of the outflow is oriented towards the southwest of it.

Moreover the velocity gradient of the CCS cannot be explained by a
pole-on outflow. An outflow in the plane of the sky, however, could sweep
the CCS material providing the velocity gradient
observed. The geometry on the plane of the sky is reinforced by the
lack of a velocity gradient of the water masers along its jet-like structure.

Considering all the available information, an outflow that lies near the
plane of the sky with a finite opening angle, can explain all the
results obtained with our observations (see Fig 1).

\begin{figure}
\centerline{\includegraphics[width=2.5in,angle=-90]{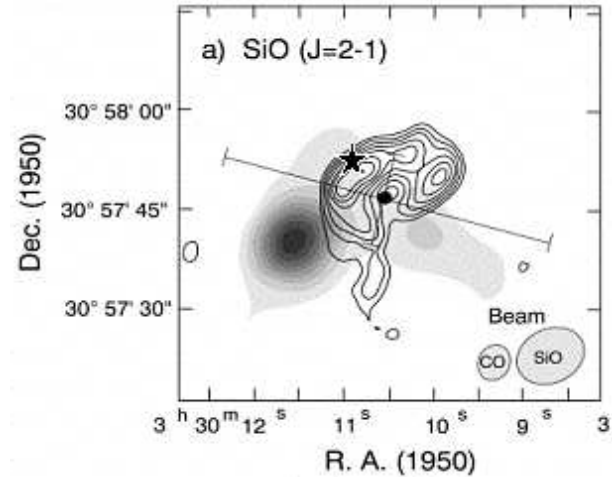}}
\caption{Contours represent the CO(1-0) integrated intensity map 
from Hirano et al. (1997) superimposed in gray scale on SiO observed by
Yamamoto et al. (1992). The black star represents the position of the 2
micron point source. The filled circle marks the position of the IRAS source.}
\end{figure}

\begin{acknowledgements}

GA, IdG, JFG and JMT acknowledge support from MCYT grant     (FED-ER funds) 
AYA2002-00376 (Spain). GA acknowledges support from Junta de Andaluc\'{\i}a (Spain). 
IdG acknowledges the support of a Calvo Rod\'es
Fellowship from the Instituto Nacional de T\'ecnica Aeroespacial and of the summer students program of the National
Radio Astronomy Observatory.

\end{acknowledgements}

\end{article}
\end{document}